\documentclass[aip,graphicx]{revtex4-1}

\usepackage{graphicx}

\draft 

\begin{document}


\title{Spectral evolution of two-dimensional kinetic plasma turbulence
in the wavenumber-frequency domain} 



\author{H.~Comi\c{s}el}
\email{h.comisel@tu-bs.de}
\affiliation{Institut f\"ur Theoretische Physik, 
Technische Universit\"at Braunschweig, Mendelssohnstr. 3, D-38016 Braunschweig, Germany}
\affiliation{Institute for Space Sciences, Atomi\c{s}tilor 409, 
P.O. Box MG-23, Bucharest-M\u{a}gurele RO-077125, Romania}

\author{D.~Verscharen}
\affiliation{Space Science Center,
University of New Hampshire, 8 College Rd., Durham, NH 03824, USA}

\author{Y.~Narita}
\affiliation{Space Research Institute, Austrian Academy of Sciences,
Schmiedlstr. 6, A-8042 Graz, Austria}

\author{U.~Motschmann}
\affiliation{Institut f\"ur Theoretische Physik, 
Technische Universit\"at Braunschweig, Mendelssohnstr. 3, D-38016 Braunschweig, Germany}
\affiliation{Deutsches Zentrum f\"ur Luft- und Raumfahrt,
Institut f\"ur Planetenforschung, Rutherfordstr. 2, D-12489 Berlin, Germany}

\date{\today}

\begin{abstract}
We present a method for studying the evolution of plasma turbulence by tracking dispersion relations in the energy spectrum in the wavenumber-frequency domain. We apply hybrid plasma simulations in a simplified two-dimensional geometry to
demonstrate our method and its applicability to plasma turbulence in the ion kinetic regime.
We identify four dispersion relations: ion-Bernstein waves, oblique whistler waves, oblique Alfv\'en/ion-cyclotron waves, and a zero-frequency mode.
The energy partition and frequency broadening are
evaluated for these modes. The method allows us to determine the evolution of decaying plasma turbulence in our restricted geometry and shows that it cascades along the dispersion relations during the early phase with an increasing broadening around the dispersion relations.
\end{abstract}


\maketitle 






Plasma turbulence is inherently nonlinear in nature \cite{howes2008,schekochihin2009}. However, the contribution and importance of linear effects is currently under debate and a key question in the Turbulent Dissipation Challenge \cite{parashar2013}. If the linear response of the system dominates over any nonlinear effects, the plasma and field fluctuations show properties similar to a superposition of linear waves. This effects manifests in the associated time scales: If the typical nonlinear time scales are much greater than the typical propagation time scales of the modes, the notion of linear wave modes is valid.  
On the other hand, if nonlinearities can induce spectral transfer between different wavenumbers on shorter time scales than any associated propagation period of waves,
the wave picture breaks down \cite{karimabadi2013,wu2013}. 
Recently, plasma turbulence research focuses on the case in which the nonlinear time scales 
and the wave-propagation time scales are equal. This situation is denoted 
critical balance \cite{goldreich1995}. Critically balanced turbulence is strong; 
however, it shows some wave properties rooted in the linear behavior of the plasma. 
One important feature of linear wave modes is their dispersion relation that associates 
frequencies with wavenumbers.

In this Letter, we present a method to quantify the power of fluctuations along given dispersion relations and thus to compare the contribution of normal modes with the contribution of turbulent structures that do not follow the normal-mode relations. This method is closely connected to the concept of the \emph{random-sweeping hypothesis}, which describes the spectral broadening associated with large-scale advection in turbulent flows \cite{kraichnan64,tennekes75,wilczek12}. 
Since dispersion relations are significantly distinguishable at kinetic scales,  
hybrid simulations provide a good tool for the detailed investigation of 
the nature of turbulent fluctuations. 
The hybrid simulations by Verscharen et al. \cite{verscharen2012} have demonstrated 
the existence of various normal modes such as whistler and Alfv\'en/ion-cyclotron 
waves as well as ion-Bernstein modes in the ion kinetic regime in a two-dimensional geometry.
We build our analysis upon similar numerical simulations of kinetic plasma turbulence
in the two-dimensional domain covering the directions parallel and perpendicular
to the mean magnetic field. We determine the wavenumber-frequency spectra
at different times of the evolution. We analyze the fluctuation energy distributed along the
dispersion relations and the spectral broadening around the dispersion relations to evaluate the accuracy of the linear-wave picture in plasma turbulence.

We perform direct numerical simulations using the hybrid code AIKEF
(Adaptive Ion-Kinetic Electron-Fluid) \cite{mueller2011}. 
The code treats ions as individual super-particles following
the characteristics of the Vlasov equation and electrons as a massless charge-neutralizing fluid. 
The code has successfully been applied to the studies of fundamental processes
of plasma turbulence, particularly in resolving ion kinetic scales, 
as well as studies of the plasma environment of solar-system bodies \cite{kriegel11,wiehle11,mueller12,verscharen12b}.

Our simulation box is aligned with the directions parallel and perpendicular
with respect to the mean magnetic field, and the boundaries are periodic in space. 
The particles obey a Maxwellian distribution at the initial time with
the width determined by the initial plasma beta, $\beta_{\rm p} = 0.05$, for the protons, and the electron beta is $\beta_{\rm e} = 0.5$.
The simulation box has a size of 250 times 250
proton inertial lengths $V_{\rm A}/\Omega_{\rm p}$
where $V_{\rm A}$ and  $\Omega_{\rm p}$ denote the Alfv\'en speed and 
the proton gyrofrequency, respectively.
The box is resolved by 2048 times 2048 cells, and each cell is filled with 400 super-particles representing the protons. All vectors are treated with three components.
The constant mean magnetic field $\vec B_0$ is aligned with the $z$-direction.
As the initial condition, we impose large-scale magnetic field fluctuations in form of a
 superposition of one thousand Alfv\'en waves with 
an isotropic wavevector distribution 
between $k_{\min}=0.05\Omega_{\mathrm p}/V_{\mathrm A}$ and $k_{\max}=0.2\Omega_{\mathrm p}/V_{\mathrm A}$
. The waves have random phases and their amplitudes follow the Kolmogorov scaling (spectral energy density proportional to $|k|^{-5/3}$) 
 in such a way that the total power of the composed wave field  equals the power of a monochromatic wave with  $\delta B=0.01 B_0$ 
.
The plasma velocity is correlated to the initial wave magnetic field according to the Alfv\'en-wave polarization relation. 
The time step in our simulation is $0.5\,\Omega_{\rm p}^{-1}$.
During the simulation, no additional energy is introduced to the system. 
This simulation setup is an extension of that by Verscharen et al. \cite{verscharen2012}. We simulate the system over a longer time so that the evolution beyond the first spectral energy transfer can be studied.

\begin{figure}[t]
\begin{center}
\includegraphics[width=8.3cm]{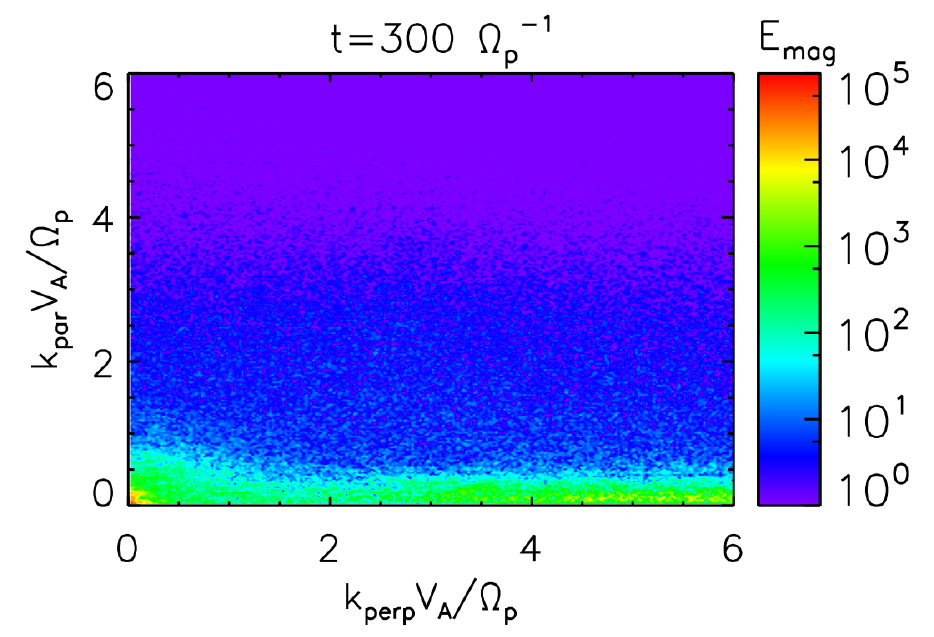}
\caption{Magnetic energy spectrum in the wavevector domain 
spanned by parallel and perpendicular components of wavevector
with respect to the mean magnetic field direction
at $t=300\,\Omega_{\mathrm p}^{-1}$.}
\label{fig:kspec}
\end{center}
\end{figure}

Fig.\,\ref{fig:kspec} shows the energy spectrum during the growth phase of the evolution
in the wavevector domain at $t=300\,\Omega_{\mathrm p}^{-1}$.
The fluctuation energy is transported to higher
wavenumbers preferentially in the direction perpendicular to the mean magnetic field, and thus the spectrum is markedly anisotropic.
Most of the fluctuation energy is limited to highly oblique angles of propagation (greater than 80 degrees) and to small parallel wavenumbers, $k_\| V_{\rm A} / \Omega_{\rm p} \le 1$.

\begin{figure*}[t]
\begin{center}
%
\includegraphics[width=16.8cm]{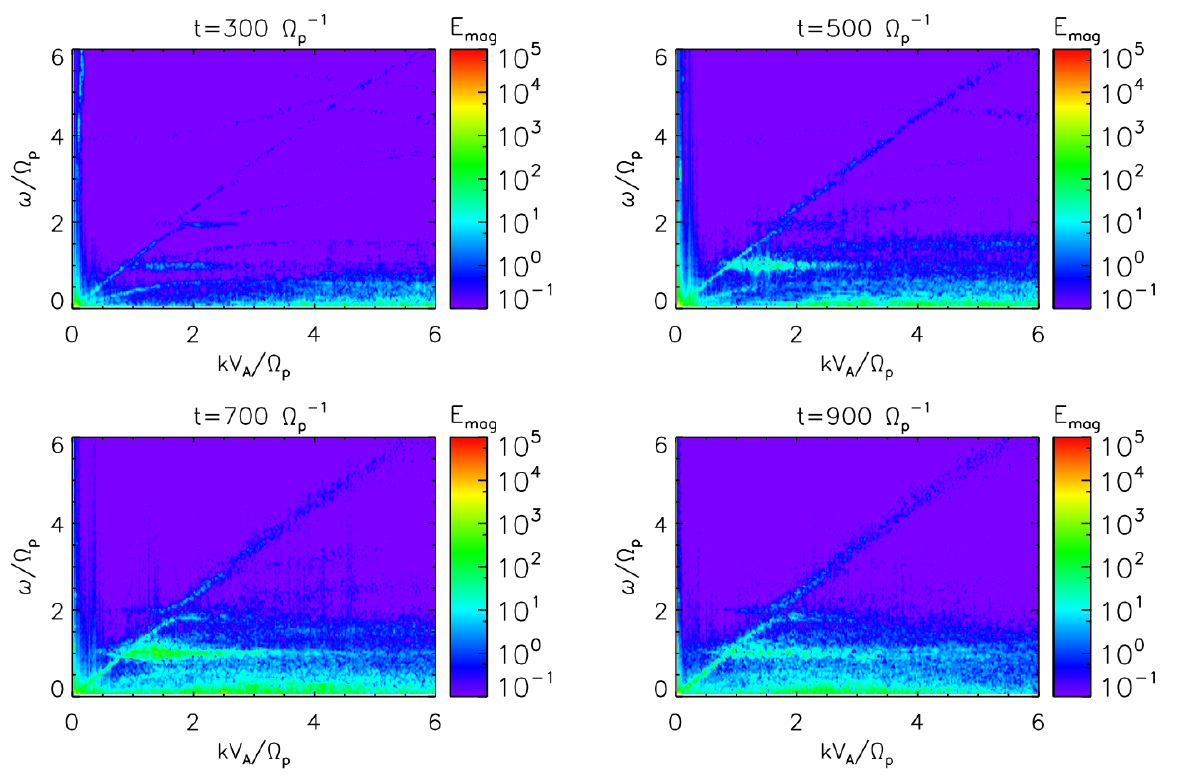}
\caption{Magnetic energy spectrum
in the perpendicular wavenumber-frequency domain
at time $t=300\,\Omega_{\rm p}^{-1}$, $t=500\,\Omega_{\rm p}^{-1}$,
$t=700\,\Omega_{\rm p}^{-1}$, and $t=900\,\Omega_{\rm p}^{-1}$.
Spectral density is normalized using the mean magnetic field $B_0$,
the Alfv\'en speed $V_{\rm A}$, and the proton gyrofrequency $\Omega_{\rm p}$.}
\label{fig:spectra}
\end{center}
\end{figure*}

In Fig. \ref{fig:spectra}, we show 
cuts of the magnetic energy spectrum for $k_{\parallel}=0$ 
at four time steps (300, 500, 700, and 900 gyroperiods) in the $\omega$-$k_{\perp}$ plane.
Fluctuation energy is obviously not transported to arbitrary wavenumbers or frequencies, but the spectrum exhibits a clear organization in wavenumbers and frequencies.
This organized pattern represents in part the dispersion relations of different linear modes
accompanied by nonlinear side-band components.
Already at $t=300\,\Omega_{\mathrm p}^{-1}$, the fluctuation energy has been very efficiently transported to higher wavenumbers at frequencies close to zero.
With increasing time, the fluctuation energy fills the
various branches seen in Fig.~\ref{fig:spectra}.

We focus on four distinct branches in the dispersion relation in the kinetic regime at $k_{\perp}V_{\mathrm A}/\Omega_{\mathrm p}\ge 1$.
For the purpose of our quantitative analysis, we identify the following modes and use the given approximations for their particular dispersion relations:
(1) a zero-frequency mode (Z) with $\omega = 0$;
(2) the first-order ion-Bernstein mode (B)  approximated by $\omega = \Omega_{\rm p}$;
(3) the oblique whistler mode (W) approximated by a second-order polynomial \cite{gary13},
\begin{equation}\label{dpwhistler}
\frac{\omega}{\Omega_{\rm p}} = a_0 + a_1 \frac{k_\perp V_{\rm A}}{\Omega_{\rm p}} + 
a_2 \left( \frac{k_\perp V_{\rm A}}{\Omega_{\rm p}} \right)^2
\end{equation}
with the coefficients $a_0 = -0.148$, $a_1 = 1.20$, and $a_2 = 0.0215$; and
(4) an oblique Alfv\'en/ion-cyclotron mode (A) approximated by a hyperbolic tangent curve,
\begin{equation}
\frac{\omega}{\Omega_{\rm p}} = b_0 + b_1 \tanh \left( \frac{k_\perp}{k_0} \right)
\end{equation}
with $b_0 = -0.0053$, $b_1 = 0.6$, and $k_0 = 1.1$.
We find that the identified Alfv\'en/ion-cyclotron wave is well reproduced by linear Vlasov theory for propagation angles about 75 degrees. This branch may be visible in our perpendicular analysis due to finite grid sampling as well as a misalignment of the
magnetic-field direction at the grid points due to the large-scale wave field.

From the theory of random sweeping \cite{wilczek12}, we assume a Gaussian power distribution in the frequency-wavevector domain according to
\begin{equation}\label{gauss}
E(\vec k,\omega)=\frac{E(\vec k)}{\sqrt{2\pi} \sigma}\exp\left(-\frac{\left(\omega-\omega_{\mathrm{dp}}(\vec k)\right)^2}{2\sigma^2}\right)
\end{equation}
where $\omega_{\mathrm{dp}}(\vec k)$ denotes the value of the frequency according to the particular dispersion relation at the given wavevector $\vec k$, and $\sigma$ characterizes the width of the distribution as a result of frequency broadening. In order to analyze the evolution of the turbulence, we measure the relative energy distributed along the respective dispersion relation and the frequency broadening around the dispersion relations by fitting Eq.~(\ref{gauss}) to the energy distribution in the $\omega$-$k$ plane.

\begin{figure}[t]
\begin{center}
\includegraphics[width=8.3cm]{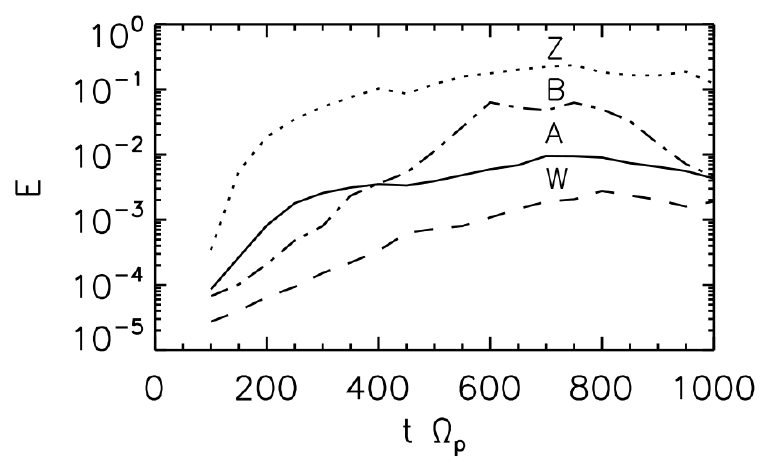}
\caption{Energy partition of different wave modes plotted as a function of time:
Z for zero-frequency mode, B for ion-Bernstein mode,
W for whistlers, and A for oblique Alfv\'en/ion-cyclotron mode.
Energy partition is normalized to the peak of the total fluctuation energy 
at $t=650\,\Omega_{\rm p}^{-1}$.}
\label{fig:partition}
\end{center}
\end{figure}

{\bf Analysis 1}. We obtain the energy partition by integrating the energy spectrum 
in the wavenumber-frequency domain along the dispersion relation of each mode, restricting ourselves only to the kinetic wavenumber range $1 \le k_\perp V_{\rm A} / \Omega_{\rm p} \le 6$. We normalize the fluctuation energy to the total fluctuation energy in the kinetic domain. Fig. \ref{fig:partition} shows the temporal evolution of energy partition among the four branches.  The zero-frequency mode grows rapidly at times between 100 and 200 gyroperiods. It gradually reaches its relative maximum at $t\approx 700\,\Omega_{\mathrm p}^{-1}$.
Afterwards, the energy partition of this mode slightly decreases while the mode itself  remains the dominant component with 25-30\,\% energy contribution
until the end of the simulation. The energy partition of the ion-Bernstein mode grows more slowly than that of the zero-frequency mode and reaches its maximum 
with about 7\,\%  at $t\approx 600\,\Omega_{\mathrm p}^{-1}$. Afterwards,
the contribution of this mode decreases down to 0.5\.\%.
The oblique Alfv\'en/ion-cyclotron mode grows also rather rapidly within the first 200 gyroperiods and reaches an energy contribution of about 0.1\,\%, thereby becoming
the mode with the second highest relative energy. In the intermediate stage after $t=400\,\Omega_{\mathrm p}^{-1}$, the ion-Bernstein mode increases in its energy partition, and the oblique Alfv\'en/ion-cyclotron mode is third in energy. This mode exhibits its maximum at $t\approx 700\,\Omega_{\mathrm p}^{-1}$ and loses relative energy only moderately during the later evolution.
The whistler mode grows most slowly at the beginning but with almost constant growth rate 
until reaching a maximum of about 0.2\,\% at $t=800\,\Omega_{\mathrm p}^{-1}$, 
when the ion-Bernstein and the oblique Alfv\'en/ion-cyclotron mode already start to decrease in energy.
After its maximum, the whistler mode slowly decreases in relative fluctuation energy.

\begin{figure}[t]
\begin{center}
\includegraphics[width=8.3cm]{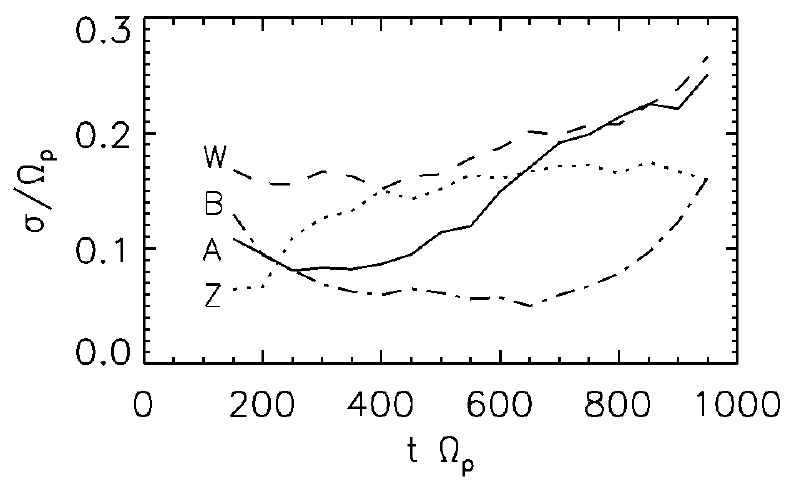}
\caption{Frequency broadening (normalized to the proton gyrofrequency) 
around the dispersion relations as a function of time.
Symbols Z, B, W, and A are the same as in Fig.\,\ref{fig:partition}.}
\label{fig:broadening}
\end{center}
\end{figure}

{\bf Analysis 2}.
In the second analysis, we determine the frequency broadening around the dispersion relation. For that purpose, we apply a fit according to Eq.~(\ref{gauss}) to the energy spectrum and determine $\sigma$ at time steps separated by 50 gyroperiods.
The values of $\sigma$ are then averaged over wavenumbers
and normalized to the proton gyrofrequency. We show the result in Fig.\,\ref{fig:broadening}.
The broadening remains below $0.3\Omega_{\mathrm p}$  for all the four modes.
While the zero-frequency mode exhibits a broadening profile
with rapid broadening first ($t\leq 300 \,\Omega_{\mathrm p}^{-1}$) and
gradual enhancement of broadening at later times,
the other modes exhibit a different type of broadening profile
with decrease first and increase later. The whistler mode slowly broadens until $t=400\,\Omega_{\mathrm p}^{-1}$. Its broadening
increases to $0.27\Omega_{\mathrm p}$ towards the end of the  simulation.
The ion-Bernstein mode and the oblique Alfv\'en/ion-cyclotron mode
show a decrease in broadening until $t=300\,\Omega_{\mathrm p}^{-1}$ and $t=600\,\Omega_{\mathrm p}^{-1}$, respectively.
At the end of the simulation, the whistler and Alfv\'en/ion-cyclotron
modes have nearly the same broadening, $0.25\Omega_{\mathrm p}-0.27\Omega_{\mathrm p}$. The zero-frequency and ion-Bernstein modes have also nearly the same broadening, $0.16\Omega_{\mathrm p}$.

Our analyses show that the wave-evolution hypothesis is a valid assumption in two-dimensional low-beta plasma turbulence in the sense that dispersion relations can be identified in the energy spectrum. 
Dispersion relations are accompanied by frequency broadening indicating side-band waves, and the effect of broadening is significant for the treated modes.
We interpret side-band-wave components as a sign of nonlinearities.
Particularly, the zero-frequency mode has a unique evolution profile among the four modes.
Its contribution in terms of energy partition is most significant with about 20\,\% of the total energy at kinetic scales. It gains its energy quickly, saturates in form of a plateau of energy partition for a long time, and hardly loses fluctuation energy afterwards. Its broadening profile shows a convex shape.
The other modes (ion-Bernstein, whistler, and oblique Alfv\'en/ion-cyclotron modes)
exhibit concave broadening profiles (broadening becomes larger after initial decrease),
even though these modes lose their fluctuation energy during the later stage of the evolution. 
We interpret this behavior as an indication that these modes lose energy not only by wave damping but also by nonlinear wave--wave couplings. Sideband waves may be excited, for example, by frequency mismatch at three-wave interaction processes \cite{gary2013}, but their existence is limited to frequencies and wavenumbers adjacent to the dispersion relation. 

The interpretation in terms of dispersion relations of the fluctuations faces several limitations in our model. First, the question about the nature of the zero-frequency mode and whether it represents coherent structures or additional wave modes cannot be answered due to the limited spectral resolution of our study.
Kinetic Alfv\'en waves may exist at very oblique propagation angles 
(85 degrees or higher), but the associated frequencies are so low that
we cannot clearly resolve them in comparison to our zero-frequency component.
Second, the whistler branch shows the typical mode-coupling behavior with the ion-Bernstein mode at $k_\perp V_{\rm A} / \Omega_{\rm p}\approx 1$ \cite{podesta12}.
In our analysis, we define whistler waves only by the continuing branch, and the mode coupling introduces inaccuracies to our prescribed dispersion relation in Eq.~(\ref{dpwhistler}). 

Our conclusions are limited to low-beta plasma turbulence. It would
be interesting to compare the energy partition and frequency broadening 
at different values of beta and for other initial plasma parameters. 

We use the constant direction of the global background magnetic field as the reference direction for our analysis. Large-scale fluctuations, however, modify the direction of the background magnetic field for small-scale fluctuations. This effect leads to part of the frequency broadening in our analysis when the scale separation is large enough. This part of the broadening results from the different definitions of a local and a global background magnetic field, which can be important for the accurate understanding of plasma turbulence \cite{boldyrev09,chen12}. 

We are aware that the energy partition among different cascade channels of turbulence is fundamentally different in two-dimensional and three-dimensional treatments \cite{tenbarge12,howes13,tenbarge13}, and that other 2D geometries (e.g., out-of-plane field) show a very different behavior from our simulation \cite{parashar09,parashar10,parashar11}. These previous works show that caution has to be exercised when quantifying different cascade channels in a limited geometry. Nevertheless, we have demonstrated that the effect of frequency broadening according to the random-sweeping hypothesis can be applied to turbulence simulations as a useful tool for the understanding of the turbulence evolution in a plasma. We propose that this analysis be applied to future fully three-dimensional simulations of plasma turbulence and complementary to solar-wind observations in order to gain insights into the relative importance of linear and nonlinear effects in solar-wind turbulence.


%
%

%

%
This work is supported by Collaborative Research Center 963, 
{\it Astrophysical Flow, Instabilities, and Turbulence} 
of the German Science Foundation as well as 
European Community's Seventh Framework Programme 
under grant agreement 313038/STORM. D.V. is supported by NASA grant NNX12AB27G.


\bibliography{draft_PoP_Comisel_20130815}


\end{document}